\documentclass[10pt]{article}
\usepackage{graphicx}

\def\Title#1{\begin{center} {\Large #1 } \end{center}}
\def\Author#1{\begin{center}{ \sc #1} \end{center}}
\def\Address#1{\begin{center}{ \it #1} \end{center}}

\newcommand\pubblock{\rightline{\begin{tabular}{l} Proceedings of the Second Annual LHCP\\ \pubnumber\\
         \pubdate  \end{tabular}}}

\newenvironment{Abstract}{\begin{quotation} \begin{center} 
             \large ABSTRACT \end{center}\bigskip 
      \begin{center}\begin{large}}{\end{large}\end{center} \end{quotation}}

\newenvironment{Presented}{\begin{quotation} \begin{center} 
             PRESENTED AT\end{center}\bigskip 
      \begin{center}\begin{large}}{\end{large}\end{center} \end{quotation}}

\def\Acknowledgements{\bigskip  \bigskip \begin{center} \begin{large}
             \bf ACKNOWLEDGEMENTS \end{large}\end{center}}

\def\beq{\begin{equation}}
\def\eeq#1{\label{#1}\end{equation}}
\def\eeqn{\end{equation}}

\def\beqa{\begin{eqnarray}}
\def\eeqa#1{\label{#1}\end{eqnarray}}
\def\eeqan{\end{eqnarray}}

\let\bar=\overbar

\def\Dslash{\not{\hbox{\kern-4pt $D$}}}
\def\dslash{\not{\hbox{\kern-2pt $\del$}}}

\def\msb{{\bar{\ssstyle M \kern -1pt S}}}

\usepackage{color}

\newcommand\pubnumber{ }

\newcommand\pubdate{\today}

\def\affiliation{
On behalf of the LHCb collaboration, \\
Department of Physics \\
University of Warwick, Coventry, CV4 7AL, U.K. }

\begin{document}

\large
\begin{titlepage}
\pubblock

\vfill
\Title{ New results in $B$ decays }
\vfill

\Author{ Mark Whitehead }
\Address{\affiliation}
\vfill
\begin{Abstract}

The latest results from studies of $b$ hadron decays to open charm final states at the LHCb experiment are presented. 
The results include measurements of $CP$ violation and the properties of $b$ baryons.

\end{Abstract}
\vfill

\begin{Presented}
The Second Annual Conference\\
 on Large Hadron Collider Physics \\
Columbia University, New York, U.S.A \\ 
June 2-7, 2014
\end{Presented}
\vfill
\end{titlepage}
\def\thefootnote{\fnsymbol{footnote}}
\setcounter{footnote}{0}
%

\normalsize 


\section{Introduction}
Decays of beauty hadrons to open charm final states are an important part of the LHCb physics programme. 
A key goal is measuring the CKM angle $\gamma$, which is the least precisely measured angle of the unitarity triangle. 
Decays such as $B^{-} \to DK^{-}$, where $D$ is $D^0$ or $\bar{D}^0$, are very important for measuring $\gamma$
by exploiting the GLW~\cite{Gronau:1990ra,Gronau:1991dp}, ADS~\cite{Atwood:1996ci} and GGSZ~\cite{Giri:2003ty} methods. 
In addition, measurements of the properties of beauty and charm hadrons can be made with excellent precision.

The LHCb detector~\cite{Alves:2008zz} is a single arm spectrometer based at the Large Hadron Collider at CERN. The detector was 
optimised to study $b$ and $c$ hadron decays. Decays of $b$ hadrons to open charm final states are predominantly 
selected by an efficient hadronic trigger. Described in this document are three analyses from LHCb, two of which 
use the full $3.0~fb^{-1}$ run one dataset and one uses $1.0~fb^{-1}$ of data collected in 2011.
 
\section{Analysis of $B^{-} \to DK^{-}$ with $D\to K^{0}_{\mathrm S}K^{\mp}\pi^{\pm}$}

The decay $B^{-} \to DK^{-}$ with $D\to K^{0}_{\mathrm S}K^{\mp}\pi^{\pm}$ provides an 
ADS-like approach to measure $\gamma$. Two categories can be defined by requiring that 
the charged kaons have the same sign (SS) or opposite sign (OS). 
A control channel $B^{-} \to D\pi^{-}$ with $D\to K^{0}_{\mathrm S}K^{\mp}\pi^{\pm}$ is used, 
which is expected to have much smaller sensitivity to $\gamma$. 
Input from CLEO-c~\cite{Insler:2012pm} is required to describe the variation of the strong phase and the 
coherence factor. These parameters are measured across the full Dalitz plot and 
in the $K^{*}(892)^{\pm}$ region, where the coherence factor and hence sensitivity to $\gamma$ is 
expected to be larger.

The full $3.0~fb^{-1}$ data sample is exploited and the $K^{0}_{S}\to \pi^{+}\pi^{-}$ final state is used.
A boosted decision tree (BDT) is used to discriminate between signal decays and combinatorial background. 
The BDT is trained using simulated signal events and a background sample from the high $B$ mass sideband.
Following the selection requirements a fit is performed to the $B$ candidate invariant mass distribution.
Signal peaks are parametrised with double Crystal Ball functions~\cite{Skwarnicki:1986xj}. The 
combinatorial background is modelled with a linear function, the partially reconstructed background by a 
non-parametric function derived from $B^{-} \to D^{0}h^{-}$ with $D^{0}\to K^{-}\pi^{+}$ data, where $h$ is 
a kaon or pion. Additionally, 
the fit to $B^{-} \to DK^{-}$ includes a cross-feed contribution from $B^{-} \to D\pi^{-}$ decays, modelled 
with a Crystal Ball function.
\begin{figure}[!htb]
 \centering
 \includegraphics[scale=0.3]{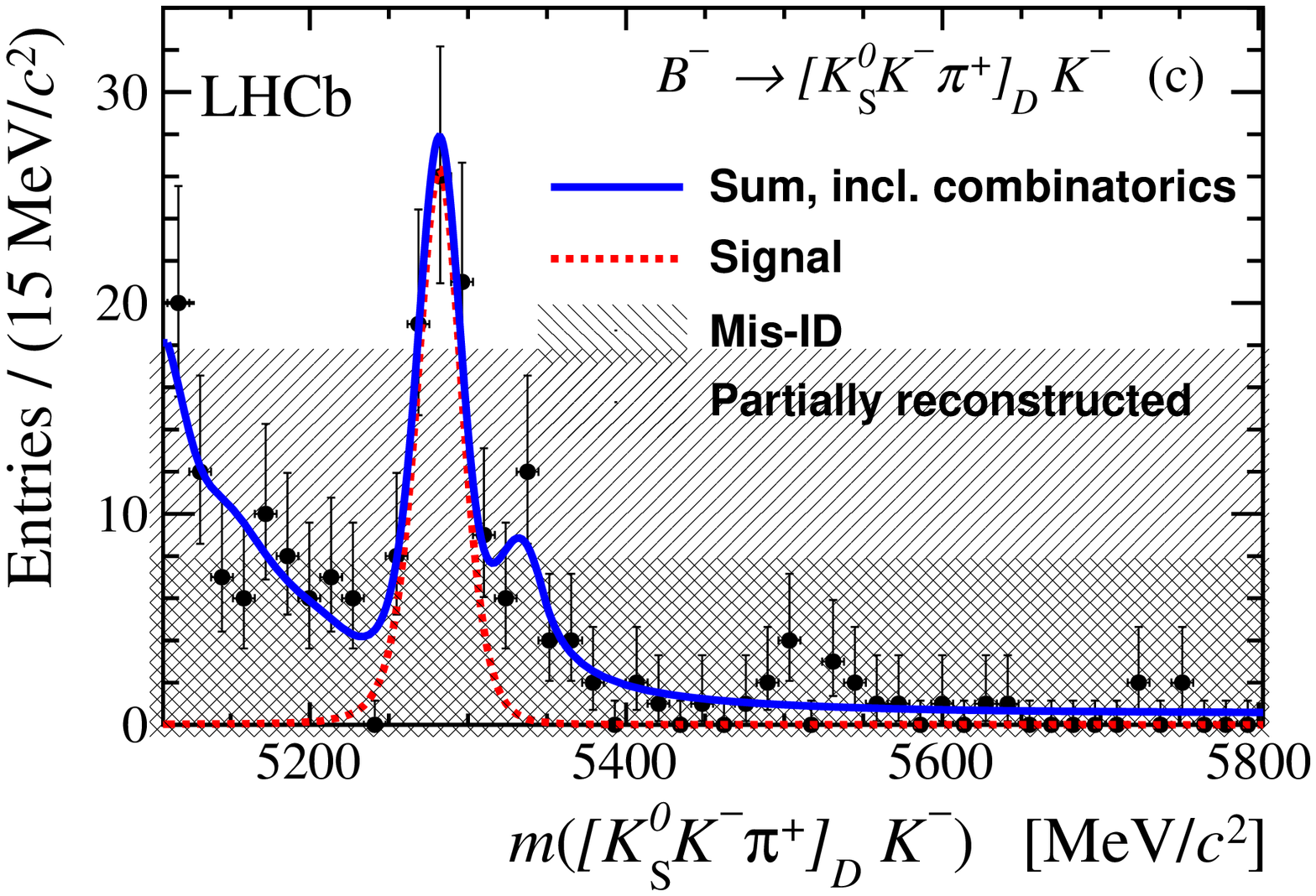}
 \includegraphics[scale=0.3]{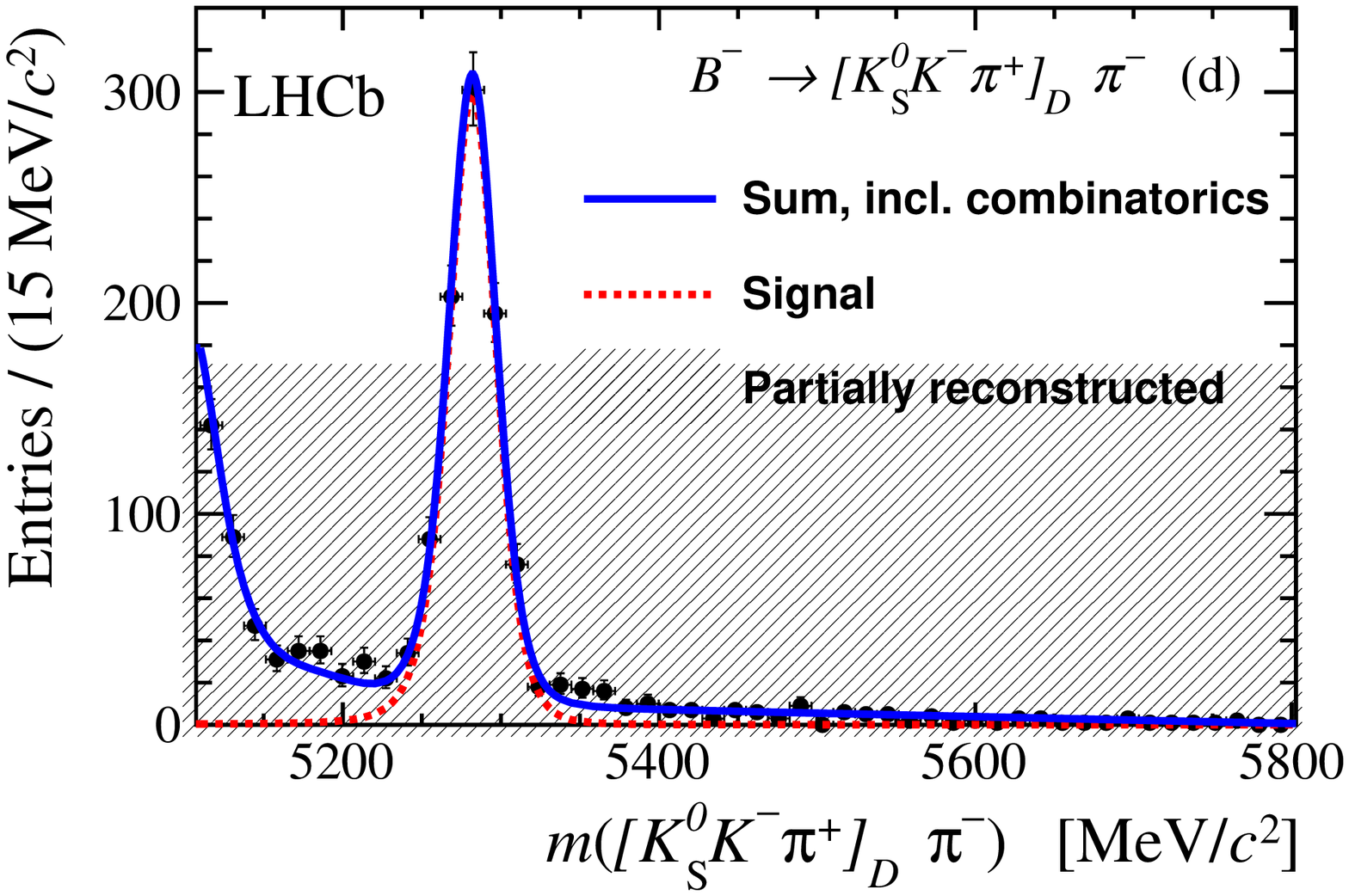}
 \caption{\small Fit to the SS $B^{-}$ candidate invariant mass distribution for (left) 
$B^{-} \to DK^{-}$ and (Right) $B^{-} \to D\pi^{-}$ decays. The legend describes the 
fit components and the black points are the data.}
 \label{fig1}
\end{figure}
An example of the $B$ candidate invariant mass fit for the SS candidates is shown in Fig.~\ref{fig1}.  
The signal yields for the various categories are summarised in Table.~\ref{tab1}. 
These are used to calculate several observables (defined in Ref~\cite{Aaij:2014dia}), three of which are effectively yield 
ratios between categories ($\mathcal{R}$). The remaining four are charge asymmetries ($\mathcal{A}$) between 
$B^{+}$ and $B^{-}$ decays. The results of these are given in Table~\ref{tab2}, where corrections due to efficiency and 
production and detection asymmetries are applied.
These observables are sensitive to both $\gamma$ and $r_{B}$, where $r_B$ is the ratio of the interfering $B^{-} \to DK^{-}$ 
amplitudes. Scans of the $\gamma - r_{B}$ parameter space is shown in Fig.~\ref{fig2}, where the $\chi^2$ probability is calculated 
at each point, taking into account statistical and systematic uncertainties. For comparison, the result of a previous LHCb 
$\gamma$ combination~\cite{Aaij:2013zfa} is shown as the black point. The results shown here are compatible with this 
and will help to improve future global fits of the angle $\gamma$. 
Note that there is increased precision on $\gamma$ in the $K^{*}(892)^{\pm}$ region, as expected.
For more details on this analysis please see Ref.~\cite{Aaij:2014dia}.
\begin{table}[!htb]
 \centering
 \caption{\small Signal yields and statistical uncertainties from the $B$ candidate invariant mass fit.}
 \label{tab1}
 \begin{tabular}{ccccc}
 \hline
  & \multicolumn{2}{c}{Whole Dalitz plot} & \multicolumn{2}{c}{$K^{*}(892)^{\pm}$ region}\\
 Mode & $DK^{\pm}$ & $D\pi^{\pm}$ & $DK^{\pm}$ & $D\pi^{\pm}$ \\
 \hline
 SS & $145\pm15$ & $1841\pm47$ & $97\pm12$ & $1365\pm38$ \\
 OS & $\phantom{0}71\pm10$ & $1267\pm37$ & $26\pm 6\phantom{0}$ & $\phantom{0}553\pm24$ \\
 \hline
 \end{tabular}
\end{table}
\begin{table}[!htb]
 \centering
 \caption{\small Value of the observables with statistical (first) and systematic (second) uncertainties.}
 \label{tab2}
 \begin{tabular}{ccc}
 \hline
 Observable & Whole Dalitz plot & $K^{*}(892)^{\pm}$ region \\
 \hline
 $\mathcal{R}_{\rm SS/OS}$ & $\phantom{-}1.528\pm0.058\pm0.025$ & $\phantom{-}\phantom{0}2.57\pm0.13\pm\phantom{0}0.06\phantom{0}$ \\
 $\mathcal{R}_{DK/D\pi\rm{, SS}}$ & $\phantom{-}0.092\pm0.009\pm0.004$ & $\phantom{-}0.048\pm0.011\pm0.003$ \\
 $\mathcal{R}_{DK/D\pi\rm{, OS}}$ & $\phantom{-}0.066\pm0.009\pm0.002$ & $\phantom{-}0.056\pm0.013\pm0.002$ \\
 $\mathcal{A}_{{\rm SS}, DK}$ & $\phantom{-}0.040\pm0.091\pm0.018$ & $\phantom{-}0.026\pm0.109\pm0.029$ \\
 $\mathcal{A}_{{\rm OS}, DK}$ & $\phantom{-}0.233\pm0.129\pm0.024$ & $\phantom{-}0.336\pm0.208\pm0.026$ \\
 $\mathcal{A}_{{\rm SS}, D\pi}$ & $-0.025\pm0.024\pm0.010$ & $-0.012\pm0.028\pm0.010$ \\
 $\mathcal{A}_{{\rm OS}, D\pi}$ & $-0.052\pm0.029\pm0.017$ & $-0.054\pm0.043\pm0.017$ \\
 \hline
 \end{tabular}
\end{table}
\begin{figure}[!htb]
 \centering
 \includegraphics[scale=0.3]{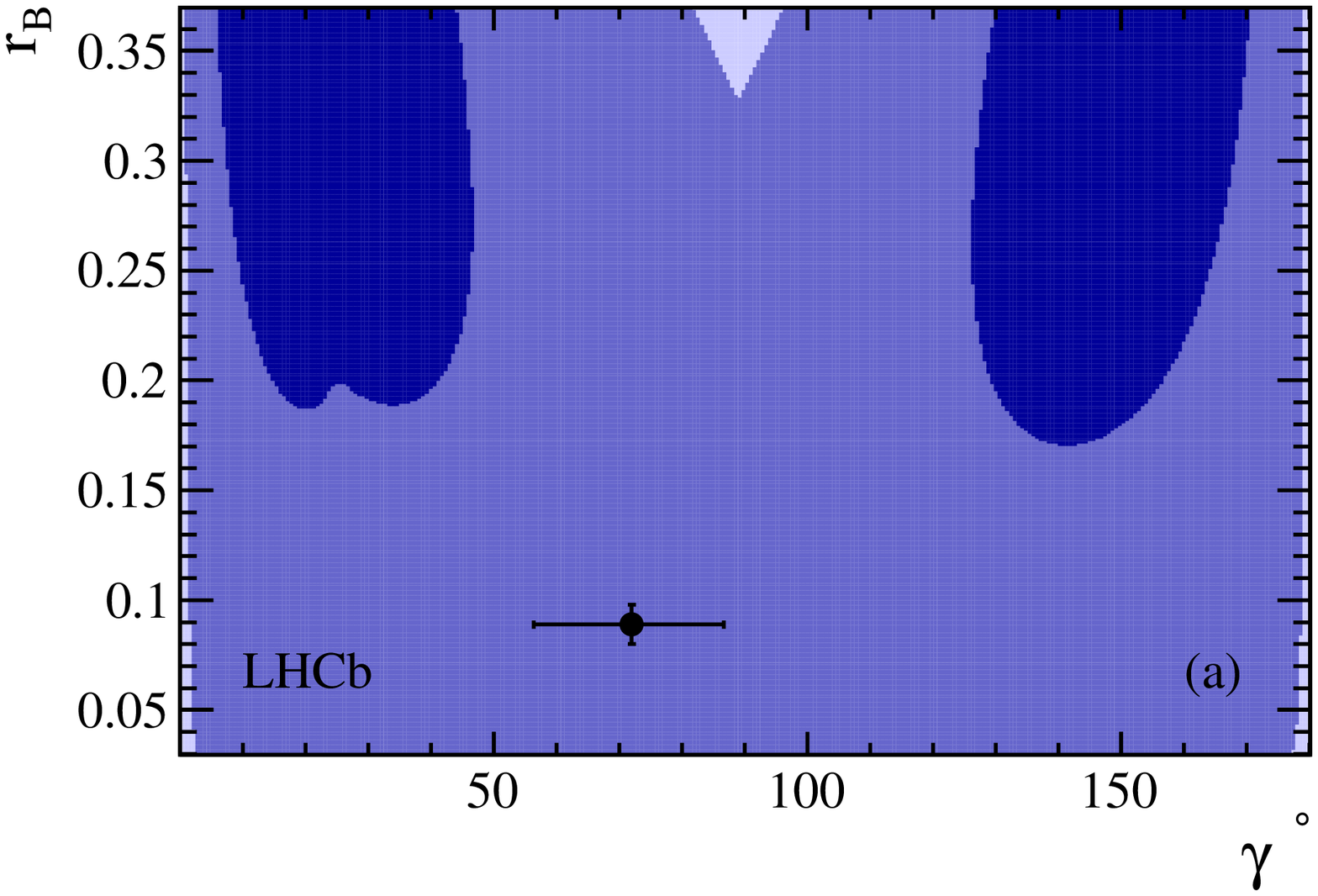}
 \includegraphics[scale=0.3]{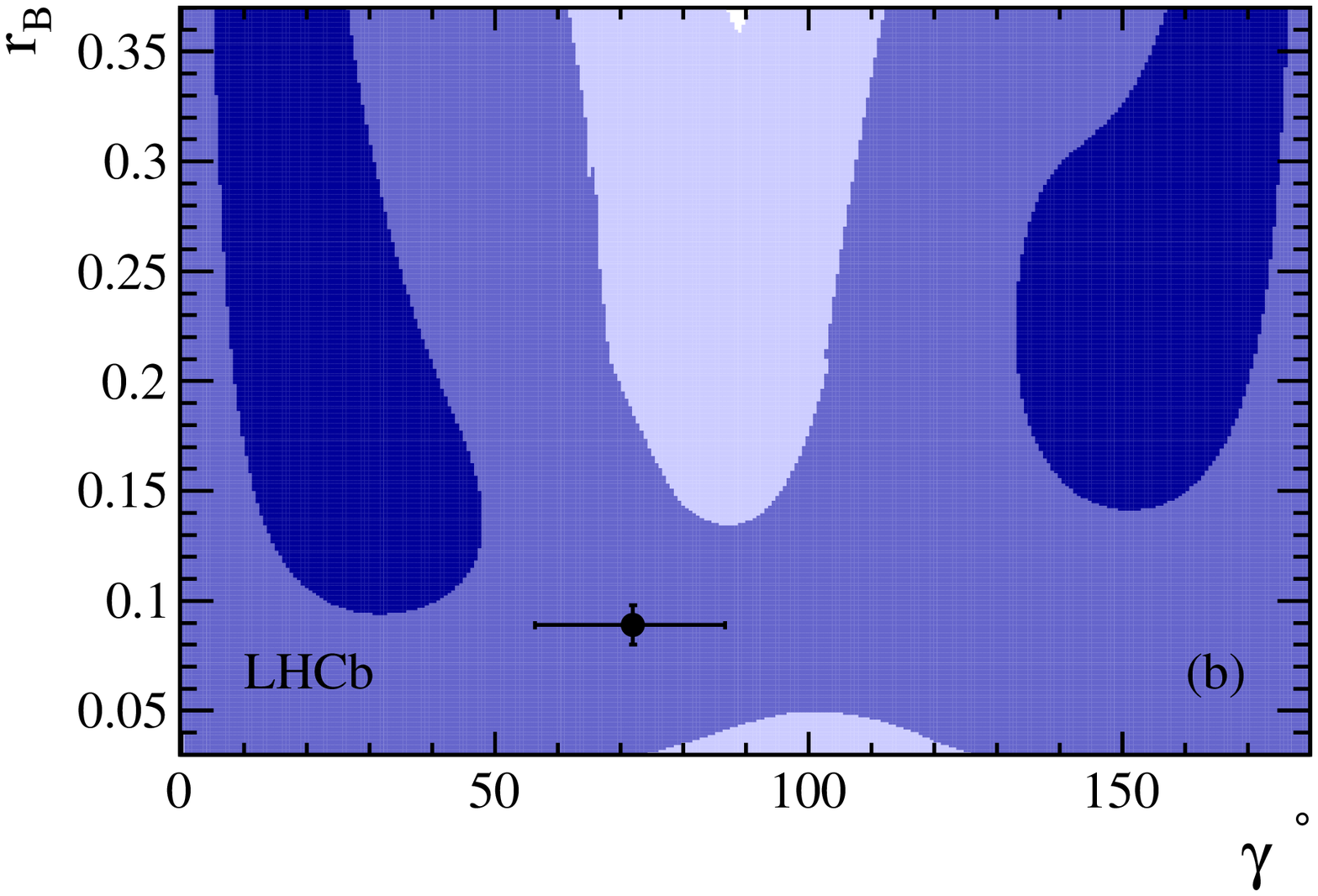}
 \caption{\small Two dimensional plot of $\gamma$ against $r_{B}$ for (left) the full Dalitz plot 
and (right) the $K^{*}(892)^{0}$ region. The contours show the (dark) $1\,\sigma$, (middle) 
$2\,\sigma$ and (light) $3\,\sigma$ of the $\chi^{2}$ probability.}
 \label{fig2}
\end{figure}

\section{Model dependent GGSZ analysis of $B^{-} \to DK^{-}$ decays}

The GGSZ method uses decays of the $D$ meson to three-body quasi-self-conjugate final states such as 
$D\to K^{0}_{\mathrm S}\pi^{-}\pi^{+}$ to make a measurement of $\gamma$. 
In this model-dependent analysis~\cite{Aaij:2014iba} an amplitude model 
from BaBar~\cite{delAmoSanchez:2010xz} is used to perform a fit to the $D\to K^{0}_{\mathrm S}\pi^{-}\pi^{+}$ Dalitz plot.
The decay $B^{-} \to D\pi^{-}$ with the same $D$ decay is used as a control mode.

\begin{figure}[!htb]
 \centering
 \includegraphics[scale=0.3]{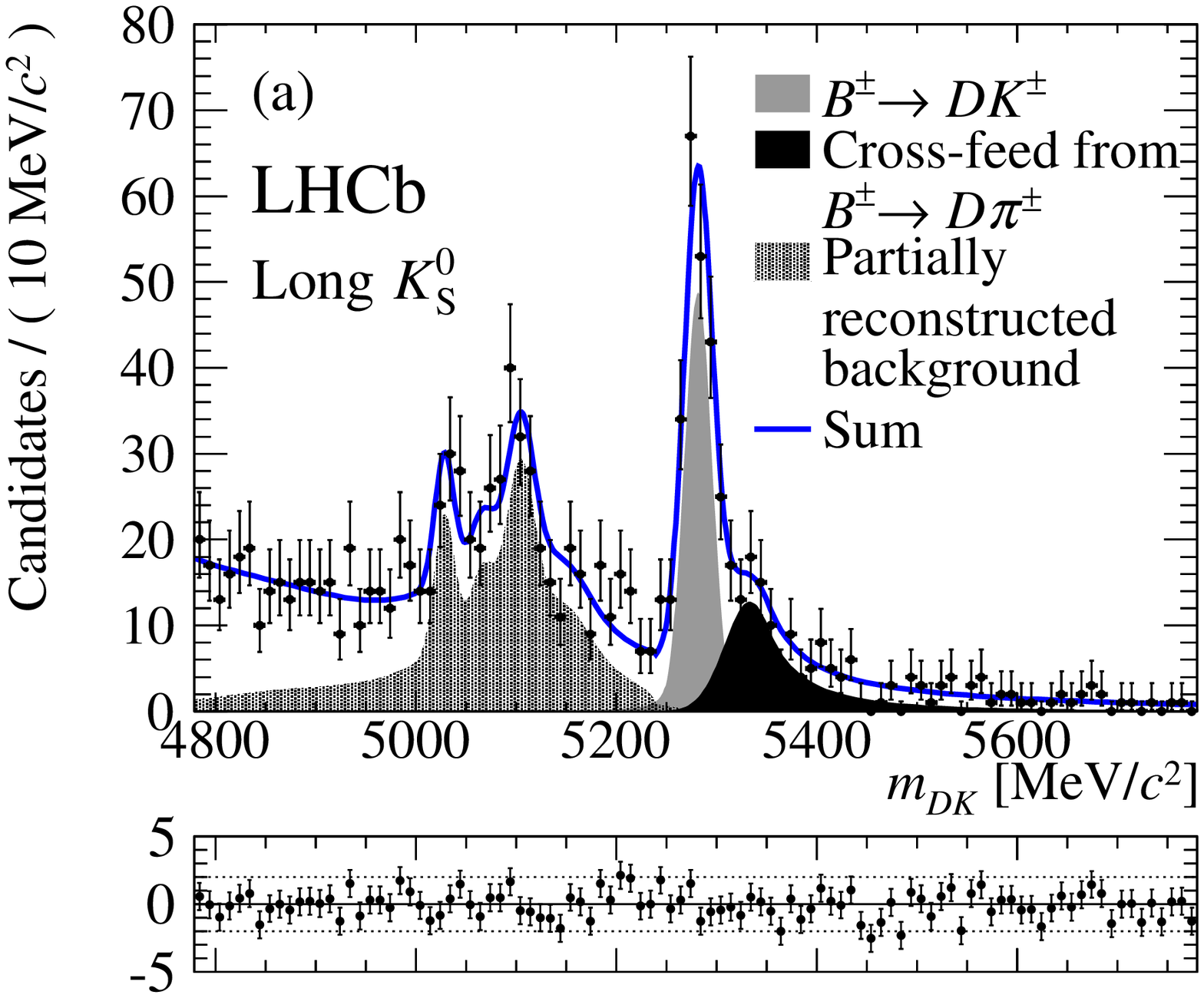}
 \includegraphics[scale=0.23]{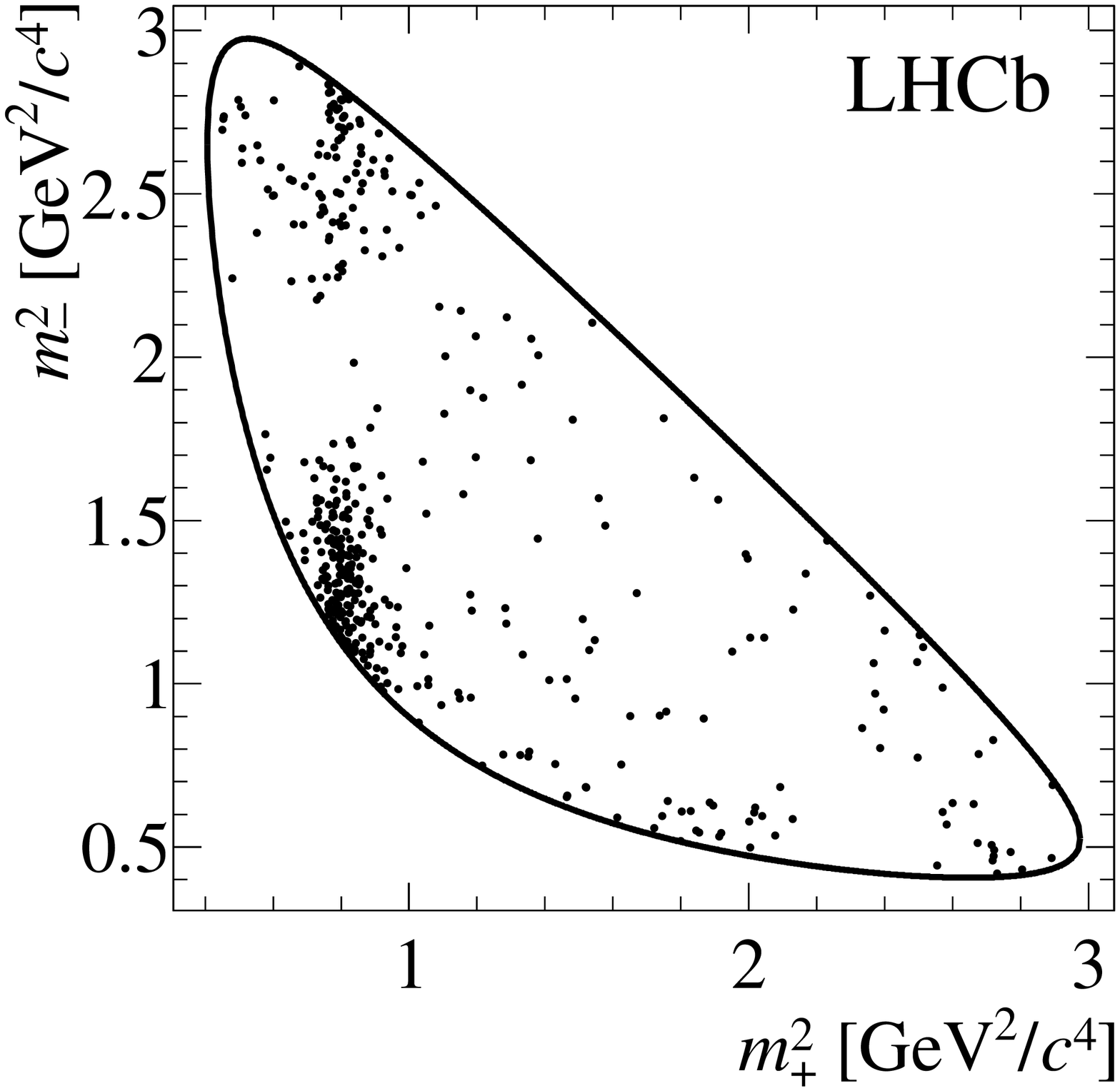}
 \caption{\small (Left) fit to the $B$ candidate invariant mass distribution for long $K_{\rm S}^{0}$ 
candidates with a normalised residual plot and (right) the $D$ Dalitz plot distribution of all $B^{+}$ candidates in the $DK$ final state.}
 \label{fig3}
\end{figure}

Candidates are selected using the $K^{0}_{\mathrm S}\to\pi^{+}\pi^{-}$ final state. The candidates are split into 
two categories depending on the $K_S^0$ reconstruction, those with both daughters having hits in the vertex locator 
are known as long candidates and those without known as downstream candidates. Following selection requirements 
on kinematic variables, and some particle identification criteria, fits to the $B$ candidate invariant mass distribution 
are performed. The mass fit is performed to determine signal and background yields which are used as an input 
to the Dalitz plot fit. The mass fits are performed simultaneously across the $DK$ and $D\pi$ final states and 
long and downstream candidates are fitted separately. An example mass fit for long candidates in the $DK$ final state 
is shown in Fig.~\ref{fig3} (left).

\begin{figure}[!htb]
 \centering
 \includegraphics[scale=0.25]{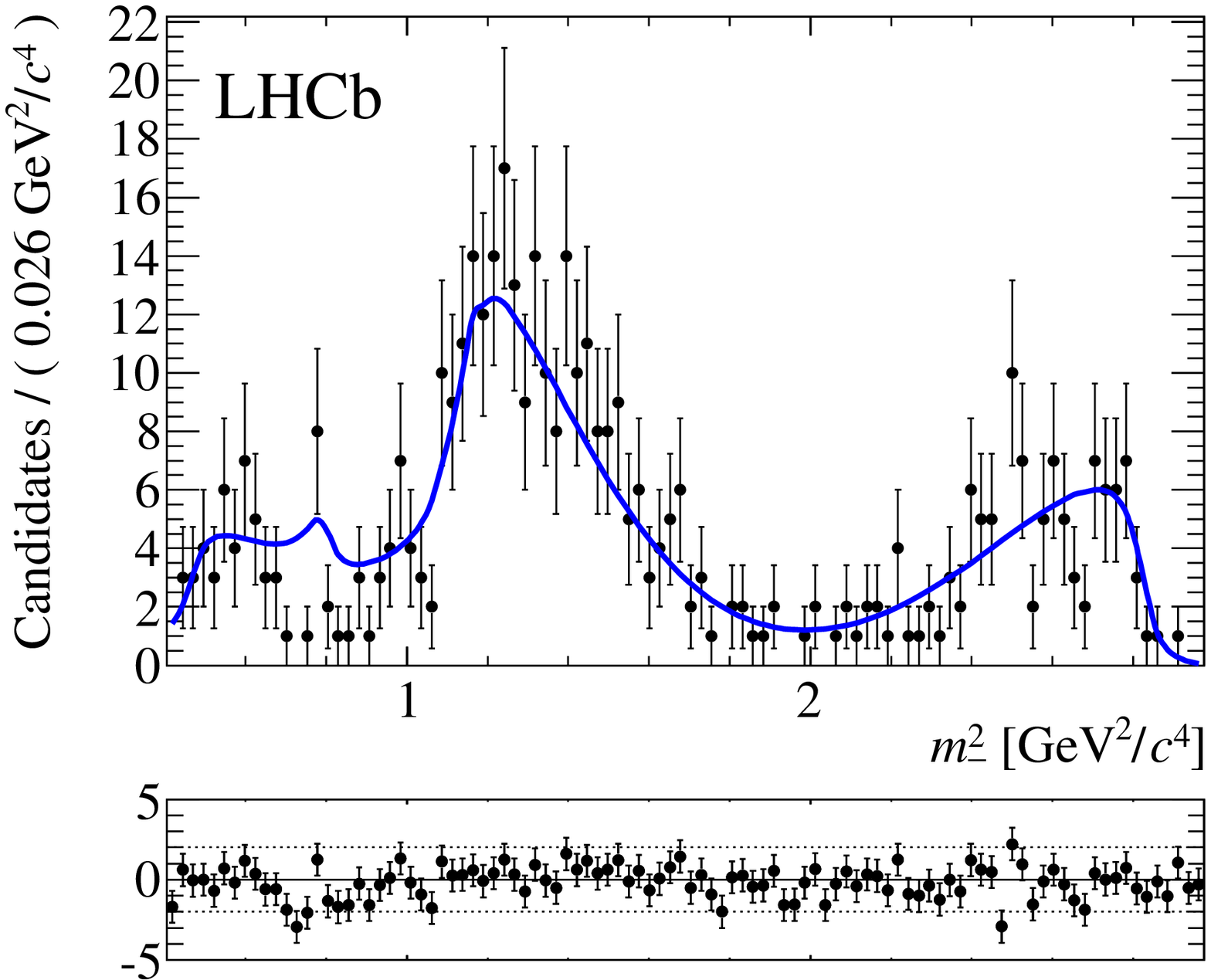}
 \includegraphics[scale=0.25]{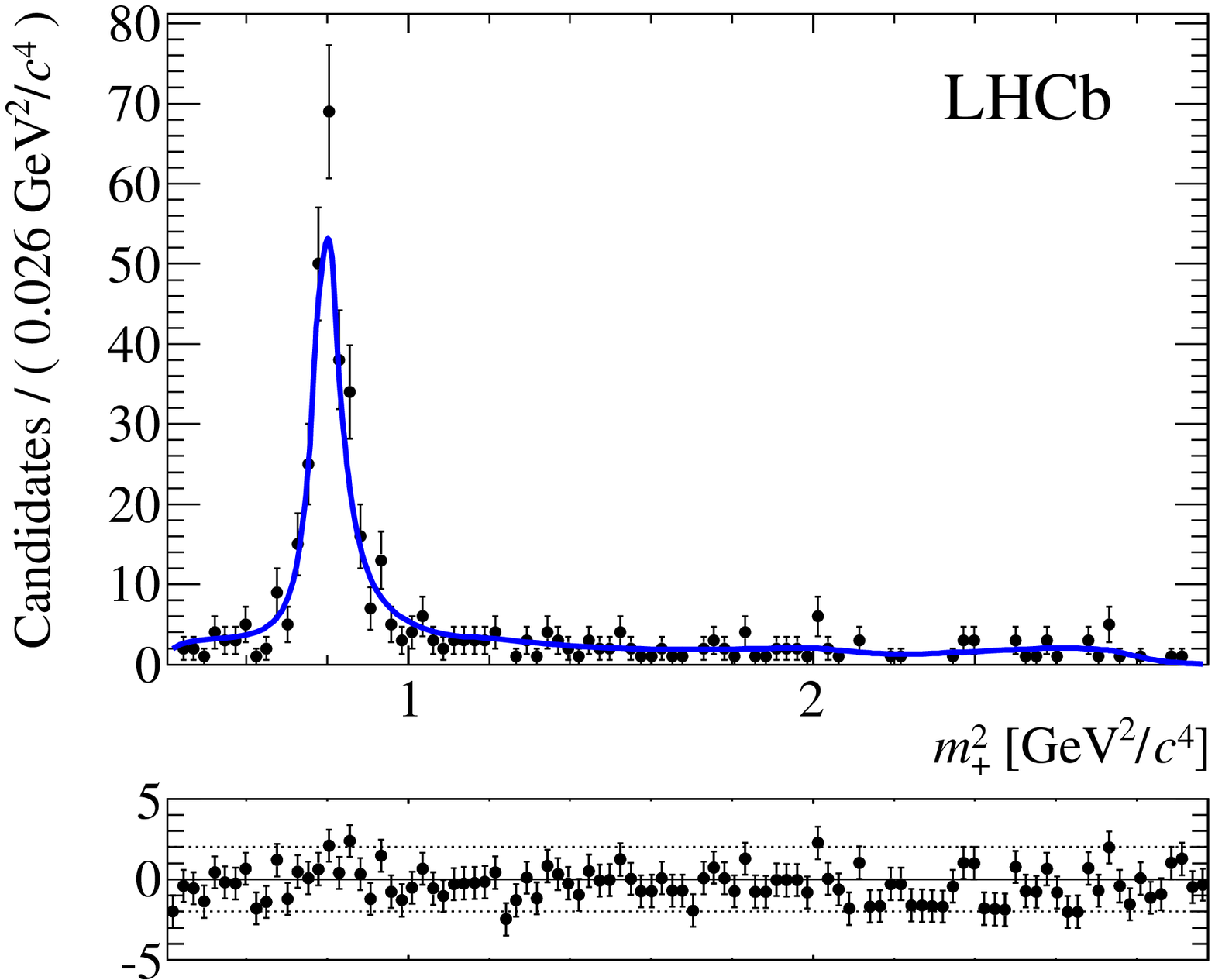}
 \includegraphics[scale=0.25]{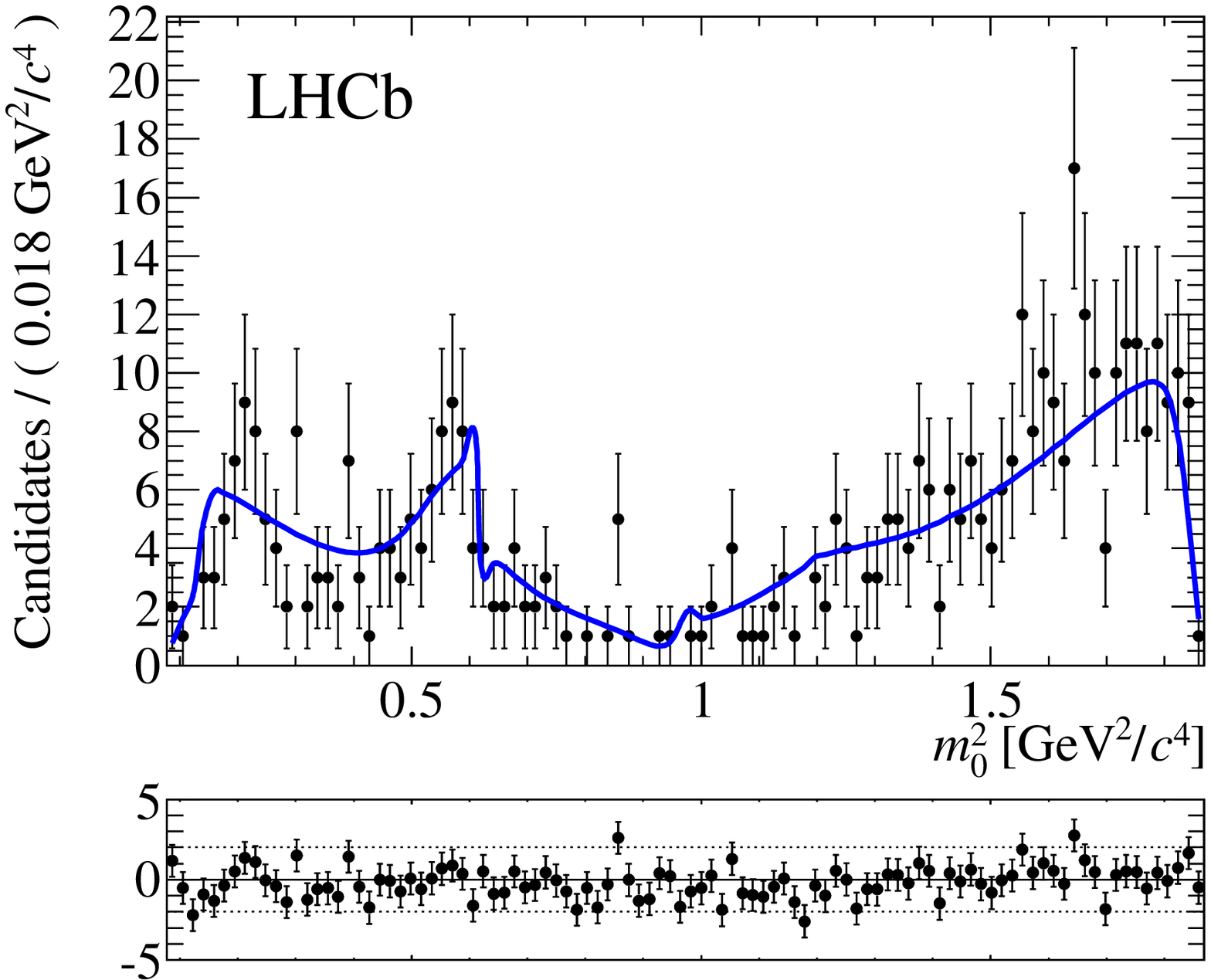}
 \caption{\small Projections of the Dalitz plot fit (blue) on to data (black) for (left) 
$m^{2}_{-}$, (middle) $m^{2}_{+}$ and (right) $m^{2}_{0}$.}
 \label{fig4}
\end{figure}

The fit to the $D$ decay Dalitz plots are performed simultaneously across long and downstream candidates, $DK$ and $D\pi$ final 
states and for $B^{+}$ and $B^{-}$ candidates. The efficiency variation, modelled as a second-order polynomial is 
determined using the $D\pi$ final state where the suppressed $B$ decay is expected to be negligible. The Dalitz plot 
distribution of $D\to K^0_S \pi^+\pi^-$ candidates from the $B^{+} \to DK^{+}$ decay is shown in Fig.~\ref{fig3} (right), 
where the $K^{*}(892)^{+}$ resonance is visible.
The fits are used to extract parameters $x_{\pm}$ and $y_{\pm}$, defined in Ref.~\cite{Aaij:2014iba}. Projections of the 
Dalitz plot fit together with the data are shown in Fig.~\ref{fig4} for $B^{+}$ candidates in the $DK$ decay mode, where 
$m_\pm=m_{K^0_S \pi^{\pm}}$ and $m_0=m_{\pi^{+}\pi^{-}}$. The values of $x_{\pm}$ and $y_{\pm}$ derived from the fit are 
\begin{eqnarray}
\centering
\nonumber x_{-} &=& +0.027 \pm 0.044 ^{\,+0.010}_{\,-0.008} \pm 0.001,\\
\nonumber y_{-} &=& +0.013 \pm 0.048 ^{\,+0.009}_{\,-0.007} \pm 0.003,\\
\nonumber x_{+} &=& -0.084 \pm 0.045 \pm 0.009 \pm 0.005,\\
\nonumber y_{+} &=& -0.032 \pm 0.004 ^{\,+0.010}_{\,-0.009} \pm 0.008,
\end{eqnarray}
where the uncertainties are statistical, systematic and from the amplitude model, respectively. From these quantities 
the values for the magnitude of the ratio of interfering $B^{\pm}$ amplitudes, the strong phase difference between them 
and $\gamma$ are found to be $r_B = 0.06 \pm 0.04$, $\delta_{B} = (115^{+41}_{-51})^{\circ}$ and $\gamma = (84^{+49}_{-42})^{\circ}$.
Neutral $D$ meson mixing has a negligible effect on each of the determined parameters. These results are consistent with
the model-independent analysis using the same data sample~\cite{Aaij:2012hu} and the world average 
values~\cite{Charles:2004jd,Bona:2007vi}. For more details on this analysis please see Ref.~\cite{Aaij:2014iba}.

\section{Measurement of the mass and lifetime of the $\Xi_{b}^{0}$ baryon}

Large samples of $b$ baryons are produced at the LHC, allowing for accurate measurements of their properties. 
Measurements of the mass and lifetime of the $\Xi_{b}^{0}$ baryon, which has quark content $bsu$, are not 
very precise~\cite{Beringer:1900zz}. The analysis~\cite{Aaij:2014esa}, based on the $3~fb^{-1}$ data sample, uses the 
$\Xi_{b}^{0} \to \Xi_{c}^{+}\pi^{-}$ decay with $\Xi_{c}^{+} \to pK^{-}\pi^{+}$ and the normalisation decay mode 
is $\Lambda_{b}^{0} \to \Lambda_{c}^{+}\pi^{-}$ with $\Lambda_{c}^{+} \to pK^{-}\pi^{+}$.

\begin{figure}[!htb]
 \centering
 \includegraphics[scale=0.3]{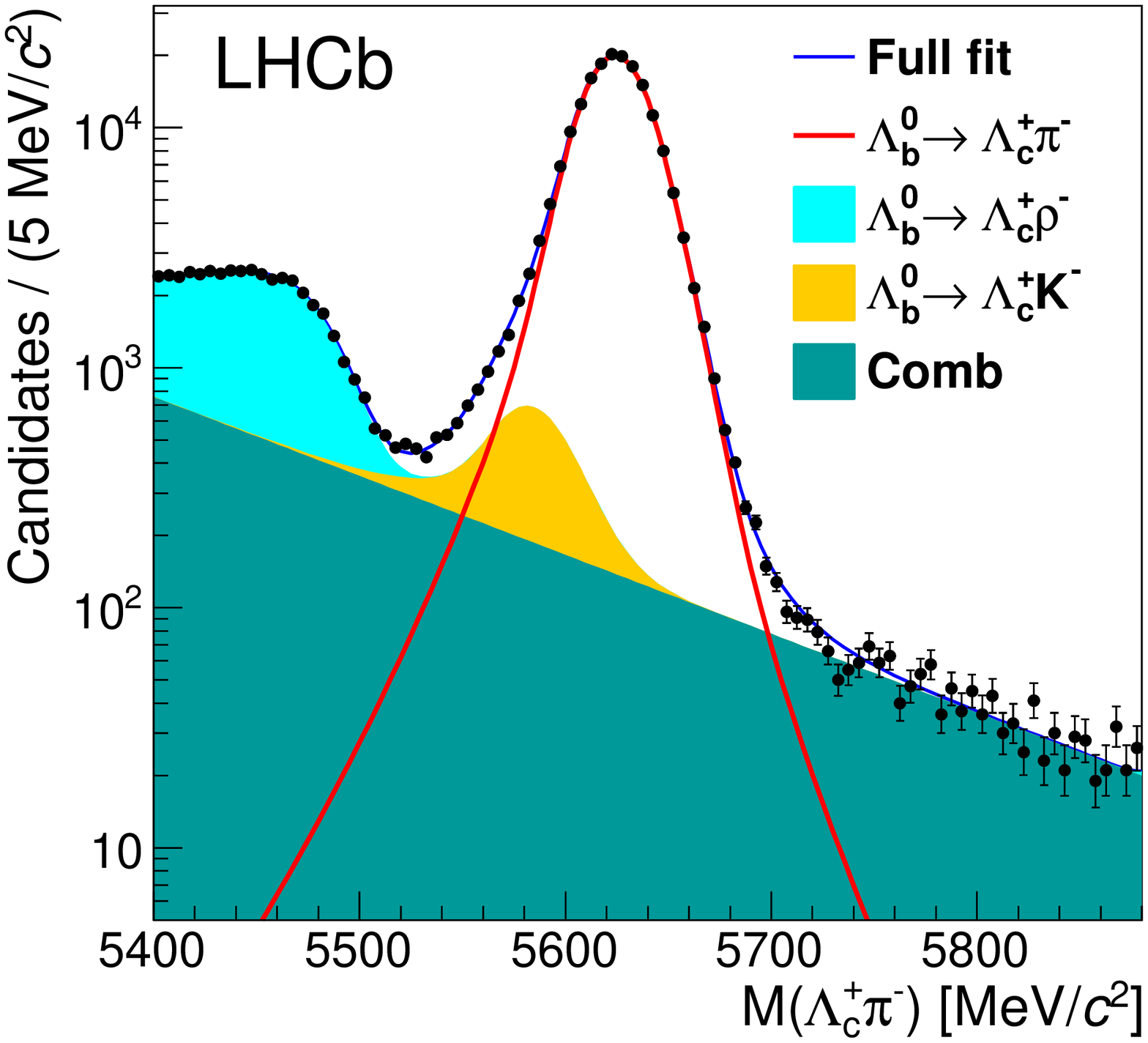}
 \includegraphics[scale=0.3]{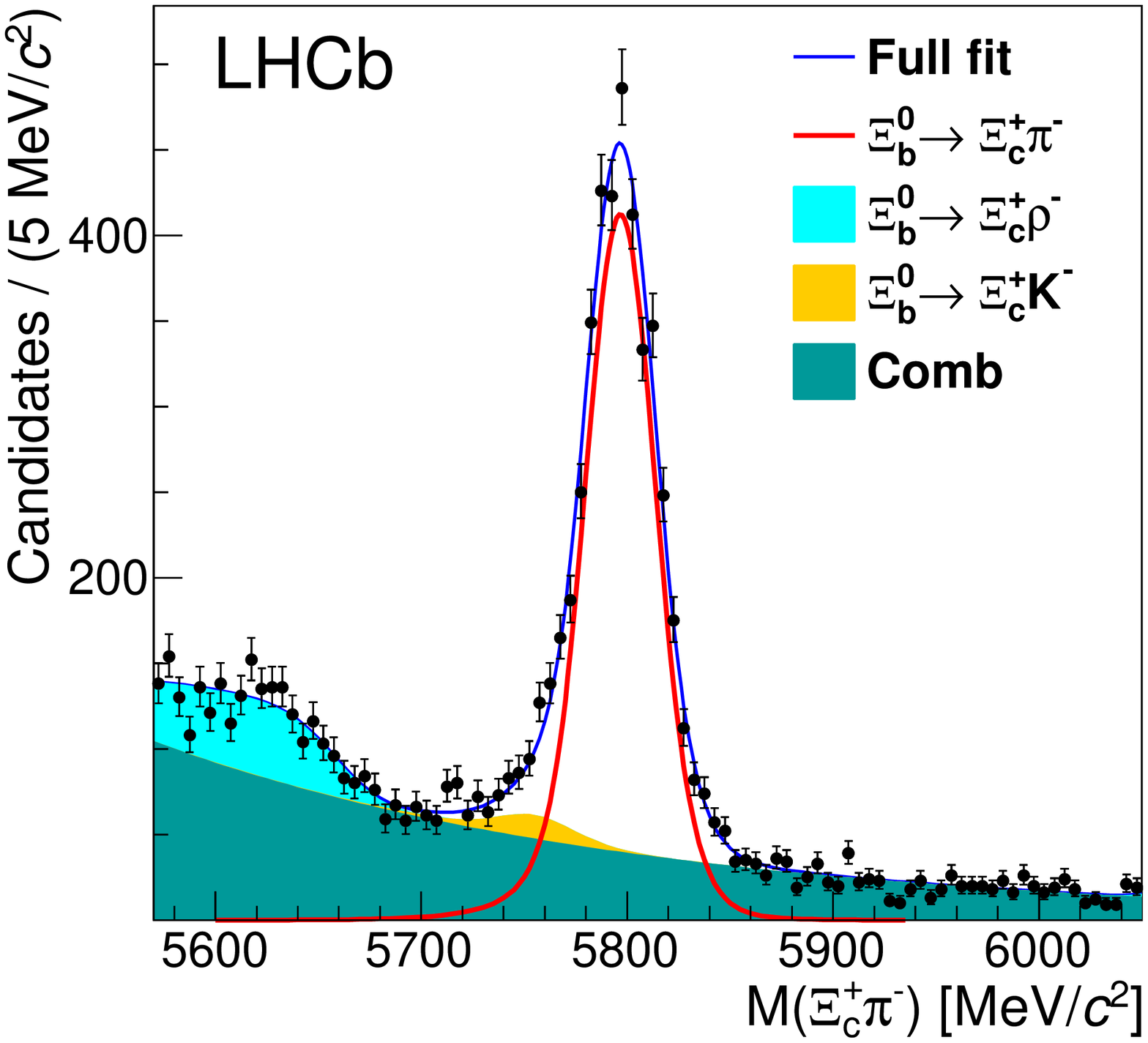}
 \caption{\small Fit to the $b$ baryon invariant mass distribution for (left) $\Lambda_{b}^{0}$ 
and (right) $\Xi_{b}^{0}$ candidates. The data are black points and the fit components 
are described in the legend.}
 \label{fig5}
\end{figure}

A BDT is used to separate signal decays from combinatorial background and is trained using signal decays from 
simulation and background from the high $b$ baryon mass sideband. A fit is performed to the $b$ baryon 
candidate invariant mass distribution to extract the signal yields. The fit is performed simultaneously 
to samples of $\Lambda_{b}^{0}$ and $\Xi_{b}^{0}$ candidates. The invariant masses of charm baryon candidates
are also fitted, to enable a mass measurement of the $\Xi_{c}^{0}$ baryon. Signals are modelled with double Crystal Ball 
functions, misidentified backgrounds are fitted with non-parametric shapes determined from simulation, an 
exponential shape is used for the combinatorial background and partially reconstructed decays are modelled by an 
ARGUS function~\cite{Albrecht:1994tb} convolved with a Gaussian. The fits are shown in Fig.~\ref{fig5} for both 
$\Lambda_{b}^{0}$ and $\Xi_{b}^{0}$ candidates. The mass differences between the two $b$ baryons and the two $c$ baryons are
determined to be
\begin{eqnarray}
\centering
\nonumber M(\Xi_{b}^{0}) - M(\Lambda_{b}^{0}) &=& 172.44 \pm 0.39 ({\rm stat}) \pm 0.17 ({\rm syst}) \,{\rm MeV}/c^{2},\\
\nonumber M(\Xi_{c}^{+}) - M(\Lambda_{c}^{+}) &=& 181.51 \pm 0.14 ({\rm stat}) \pm 0.10 ({\rm syst}) \,{\rm MeV}/c^{2}.
\end{eqnarray}
Using the world average values for $ M(\Lambda_{b}^{0})$ and $M(\Lambda_{c}^{+})$ these become
\begin{eqnarray}
\centering
\nonumber M(\Xi_{b}^{0}) &=& 5791.80 \pm 0.39 ({\rm stat}) \pm 0.17 ({\rm syst}) \pm 0.26 (M(\Lambda_{b}^{0})) \,{\rm MeV}/c^{2},\\
\nonumber M(\Xi_{c}^{+}) &=& 2467.97 \pm 0.14 ({\rm stat}) \pm 0.10 ({\rm syst}) \pm 0.14 (M(\Lambda_{c}^{+})) \,{\rm MeV}/c^{2}.
\end{eqnarray}

The ratio of lifetimes of the $\Xi_{b}^{0}$ and $\Lambda_{b}^{0}$ is measured by performing a fit 
with an exponential function to the efficiency corrected yield ratio in bins of decay time. This yields 
a ratio 
\begin{equation}
\centering
\nonumber \frac{\tau_{\Xi_{b}^{0}}}{\tau_{\Lambda_{b}^{0}}} = 1.006 \pm 0.018 ({\rm stat}) \pm 0.010 ({\rm syst}) \,{\rm ps},
\end{equation}
which can be further reduced using the known value of $\tau_{\Lambda_{b}^{0}}$~\cite{Beringer:1900zz} to give
\begin{equation}
\centering
\nonumber \tau_{\Xi_{b}^{0}} = 1.477 \pm 0.026 ({\rm stat}) \pm 0.014 ({\rm syst}) \pm 0.013 (\tau_{\Lambda_{b}^{0}}) \,{\rm ps}.
\end{equation}
These measurements are the most precise to date, and are consistent with predictions. 
This analysis demonstrates the potential of the LHCb dataset for doing high precision and high statistics 
studies of more $b$ baryons in the future. For further details of this analysis please see Ref.~\cite{Aaij:2014esa}.

\section{Conclusions}
The results of three recent LHCb analyses have been presented. They include two measurements that provide sensitivity to $\gamma$. 
The other result shows excellent progress in measuring the properties of the $\Xi_{b}^{0}$ baryon. These measurements give 
an indication of the power of the LHCb experiment to produce exciting results in the future using decays of $b$ hadrons 
to open charm final states.
 
\Acknowledgements
I thank the members of the LHCb collaboration for their help in preparing the talk and 
this document. Work supported by the European Research Council under FP7.

\end{document}